\documentclass[%
 reprint,
 superscriptaddress,
 amsmath,amssymb,
 aps,
]{revtex4-1}

\usepackage{graphicx}
\usepackage{dcolumn}
\usepackage{bm}
\usepackage{xcolor}
\usepackage{url}

\usepackage{subfigure}
\usepackage{multirow}
\usepackage{ulem}
\usepackage{cancel}

\begin{document}


\title{An Analytical Solution of the Balitsky-Kovchegov Equation with the Homogeneous Balance Method}

\author{Xiaopeng Wang}
\affiliation{Institute of Modern Physics, Chinese Academy of Sciences, Lanzhou 730000, China}
\affiliation{University of Chinese Academy of Sciences, Beijing 100049, China}
\affiliation{Lanzhou University, Lanzhou 730000, China}

\author{Yirui Yang}\email{These authors contributed equally: Xiaopeng Wang and Yirui Yang. }
\affiliation{Lanzhou University, Lanzhou 730000, China}

\author{Wei Kou}
\affiliation{Institute of Modern Physics, Chinese Academy of Sciences, Lanzhou 730000, China}
\affiliation{University of Chinese Academy of Sciences, Beijing 100049, China}

\author{Rong Wang}
\email{rwang@impcas.ac.cn}
\affiliation{Institute of Modern Physics, Chinese Academy of Sciences, Lanzhou 730000, China}
\affiliation{University of Chinese Academy of Sciences, Beijing 100049, China}

\author{Xurong Chen}
\email{xchen@impcas.ac.cn}
\affiliation{Institute of Modern Physics, Chinese Academy of Sciences, Lanzhou 730000, China}
\affiliation{University of Chinese Academy of Sciences, Beijing 100049, China}
\affiliation{Guangdong Provincial Key Laboratory of Nuclear Science, Institute of Quantum Matter, South China Normal University, Guangzhou 510006, China}


\date{\today}

\begin{abstract}
Nonlinear QCD evolution equations are essential tools in understanding the saturation
of partons at small Bjorken $x_{\rm B}$, as they are supposed to restore an upper bound of unitarity
for the cross section of high energy scattering.
In this paper, we present an analytical solution of Balitsky-Kovchegov (BK) equation using the homogeneous balance method.
The obtained analytical solution is similar to the solution of a traveling wave.
By matching the gluon distribution in the dilute region which is determined from the global analysis
of experimental data (CT14 analysis), we get a definitive solution of the dipole-proton forward scattering amplitude
in the momentum space. Based on the acquired scattering amplitude and the behavior of geometric scaling,
we present also a new estimated saturation scale $Q_s^2(x)$.
\end{abstract}

\maketitle


\section{Introduction}
\label{sec:intro}

The powerful and sharpest way to resolve the proton structure is by
the lepton deep inelastic scattering (DIS) off the proton at high energy.
The current experiments show that the gluon distribution grows rapidly toward smaller $x_{\rm B}$.
The fast growth of gluons is described by the well-established
Balitsky-Fadin-Kuraev-Lipatov (BFKL) equation \cite{Lipatov:1976zz,Kuraev:1977fs,Balitsky:1978ic},
which is derived with perturbative quantum chromodynamics (pQCD) by resumming
the leading logarithmic contribution (${\rm ln}(1/x)$) to the scattering off the proton.
However, under the assumption of the Regge-like growth of the gluon distribution as $x_{\rm B}$
decrease, the unitarity limit \cite{Froissart:1961ux,Martin:1962rt} of the $\gamma^{*}$-p cross section eventually is broken.
To restore the unitarity upper bound in QCD theory is an interesting physics,
which has been discussed for decades. The first idea is to include the parton-parton recombination process
\cite{Gribov:1981ac,Gribov:1984tu,Mueller:1985wy,Mueller:1989st,Zhu:1999ht}
which will cease the growth of gluon distribution when the scattering happens in the high density region.

An interesting and successful theory which permits the gluon saturation is
Jalilian-Marian-Iancu-McLerran-Weigert-Leonidov-Kovner (JIMWLK) equation \cite{Balitsky:1995ub,JalilianMarian:1997jx,Iancu:2000hn,Weigert:2000gi},
in which the nonlinear correction of the strong field is considered with Wilson renormalization group approach.
The saturation state predicted by JIMWLK equation is called color glass condensate (CGC) \cite{Iancu:2000hn,Iancu:2001ad,JalilianMarian:2005jf,Gelis:2010nm}.
However, JIMWLK is a complex partial derivative functional equation and it is hard for one to solve.
Another nonlinear evolution equation is Balitsky-Kovchegov (BK) equation \cite{Balitsky:1997mk,Kovchegov:1999yj,Kovchegov:1999ua,Balitsky:2001re},
in which the correction due to the resuming of the fan diagrams (two Pomerons merge into one Pomeron)
are added to the standard BFKL evolution process.
In JIMWLK equation, the quantum fluctuation is added to the evolution for the strong gluon field at small-$x$,
while in BK equation, the quantum corrections from resuming multiple rescatterings is implemented for the dipole forward amplitude.
Both equations are derived in the framework of the quantum evolution process.
In a simple view, BK equation is regarded as the mean-field approximation of JIMWLK equation.
The non-saturating regime and the saturating regime are well connected by BK equation,
and the unitarization of the high energy hadron scattering can be realized as well.
Moreover, BK equation can be solved easily compared to JIMWLK equation,
at least numerically \cite{Armesto:2001fa,GolecBiernat:2001if,Enberg:2005zj,Marquet:2005zf}.
It is an integro-differential equation, which can be transformed into a partial derivative equation in the momentum space.
Recently some analytical solutions of BK equation are proposed \cite{Munier:2003vc,Munier:2003sj,Munier:2004xu,Xiang:2017fjr,Xiang:2019kre}
from different approaches with some minor approximations.
These solutions provide some interesting insights on the nonlinear corrections to
the BFKL evolution and the phenomenological applications at high energy hadron scattering.
The applications of BK equation in explaining the experimental results
are important for us to understand the small-$x$ physics and the parton saturation.

The relation between the BK equation and the Fisher-Kolmogorov-Petrovsky-Piscounov (FKPP) equation
has been found \cite{Munier:2003vc,Munier:2003sj,Munier:2004xu}.
The FKPP equation is a famous reaction-diffusion equation in statistical physics \cite{Iancu:2004es,Munier:2014bba,Mueller:2018zwx,Enberg:2005uv},
which can be simulated easily using Monte-Carlo technique. Analytically,
the geometric scaling \cite{Stasto:2000er} observed at small-$x$ can be explained with
the traveling wave solution of FKPP equation.
It is shown in a pioneering work that the transition to the parton saturation region
in high energy QCD is identical to the formation of the front of a traveling wave \cite{Munier:2003vc,Munier:2003sj}.
Successful applications have been made in explaining the DIS data at HERA collider,
with a parametrization of the travel wave solution \cite{deSantanaAmaral:2007zzb,deSantanaAmaral:2006fe}.
In this work we present a general solution of the FKPP equation from the homogeneous balance method.
With some transformations and by matching to the gluon distribution in the non-saturating region,
we provide a definitive solution of the BK equation.

The organization of the paper is as follows.
The BK equation and the FKPP equation are reviewed in Sec. \ref{sec:equations}.
The analytical solutions of FKPP equation are introduced in Sec. \ref{sec:FKPPSolution}.
The definitive and analytical solutions of BK equation are shown in Sec. \ref{sec:BKSolution},
for the physical forward dipole-proton scattering amplitude in the momentum space.
At the end, some discussions and a summary are given in Sec. \ref{sec:summary}.

\section{BK equation and FKPP equation}
\label{sec:equations}

In the dipole picture, the DIS cross section of virtual photon is factorized into the photon wave
function $\Psi$ splitting into a color dipole ${\rm q \bar{q}}$ and the forward dipole-proton scattering amplitude
$N$ \cite{Mueller:1993rr,Mueller:1994jq,Mueller:1994gb}.
In the leading logarithm approximation, the cross section is written as \cite{Mueller:1994jq},
\begin{equation}
\begin{split}
\sigma^{\rm \gamma^* p} = \int_{0}^{\infty} rdr \int_{0}^{1} dz |\Psi(z,rQ)|^2 N(r,Y),
\end{split}
\label{eq:dipole-formula}
\end{equation}
in which $z$ is the longitudinal momentum fraction carried by the quark of the virtual photon,
$r$ is the size of the dipole, and $Y={\rm ln}(1/x)$ is the total rapidity.

The BK equation is a QCD evolution equation for describing the rapidity-dependence
of the imaginary part of the scattering between a dipole and the proton.
For the scattering amplitude ${\cal N}(Y, k)$ in the momentum space,
BK equation is given by \cite{Kovchegov:1999ua},
\begin{equation}
\begin{split}
\frac{\partial {\cal N}(k,Y)}{\partial Y} = \frac{\alpha_{\rm s} N_{\rm c}}{\pi} \chi\left(-\frac{\partial}{\partial {\rm ln}k^2}\right){\cal N}(k,Y)\\
- \frac{\alpha_{\rm s} N_{\rm c}}{\pi} {\cal N}^2(k,Y),
\end{split}
\label{eq:BK-equation}
\end{equation}
where
\begin{equation}
\begin{split}
\chi(\lambda) = \psi(1) - \frac{1}{2} \psi\left(1-\frac{\lambda}{2}\right) - \frac{1}{2} \psi\left(\frac{\lambda}{2}\right),
\end{split}
\label{eq:BFKL-kernel}
\end{equation}
is the BFKL kernel with $\psi(\lambda)=\Gamma^{\prime}(\lambda)/\Gamma(\lambda)$.
In $\psi(\lambda)$, $\lambda=-\partial / \partial{\rm ln}k^2$ is a differential operator
acting on ${\cal N}(Y,k)$,
which is a way of writing the integral kernel of the BK equation in coordinate space
as the differential operator in the momentum space after a Fourier transform.
Note that the original BK equation is an integro-differential equation
in the coordinate space.

For a commonly used approximation and defining $L={\rm ln}(k^2/k_0^2)$, S. Munier and R. Peschanski suggest
an expansion of BFKL kernel to the second order around $\lambda=1/2$ \cite{Munier:2003vc},
\begin{equation}
\begin{split}
\bar{\chi}\left(-\frac{\partial}{\partial L}\right) = \chi\left(\frac{1}{2}\right)
+ \frac{\chi^{\prime\prime}(\frac{1}{2})}{2}\left(\frac{\partial}{\partial L}+\frac{1}{2}\right)^2.
\end{split}
\label{eq:kernel-expansion}
\end{equation}
With the above expansion and the following transformations of the variables as,
\begin{equation}
\begin{split}
s = (1-\gamma)\left( L+\frac{\bar{\alpha}\chi^{\prime\prime}(\frac{1}{2})}{2} Y \right),\\
t = \frac{\bar{\alpha} \chi^{\prime\prime}(\frac{1}{2})}{2} (1-\gamma)^2 Y,\\
u(s,t) = \frac{2}{\chi^{\prime\prime}(\frac{1}{2})(1-\gamma)^2}\times \\
 {\cal N}\left(\frac{s}{1-\gamma} - \frac{t}{(1-\gamma)^2}, \frac{2t}{\bar{\alpha}\chi^{\prime\prime}(\frac{1}{2})(1-\gamma)^2}\right),\\
\bar{\alpha} = \frac{\alpha_{\rm s} N_{\rm c}}{\pi},\\
\gamma = 1 - \frac{1}{2} \sqrt{1+8\chi\left(\frac{1}{2}\right) \bigg/ \chi^{\prime\prime}\left(\frac{1}{2}\right)},
\end{split}
\label{eq:variable-transform}
\end{equation}
the BK equation turns into the FKPP equation for $u(s,t)$, which is written as \cite{Munier:2003vc},
\begin{equation}
\begin{split}
\partial_t u(s,t) = \partial_s^2 u(s,t) + u(s,t) - u^2(s,t).
\end{split}
\label{eq:FKPP}
\end{equation}
Thus, seeking for the analytical solution of BK equation becomes a problem of
finding the analytical solution of FKPP equation.
The FKPP equation is a famous nonlinear reaction-diffusion equation
in statistical physics, which has already been studied with some
systematical methods.

\section{Solutions of FKPP equation with Homogeneous Balance Method}
\label{sec:FKPPSolution}

\begin{figure}[htbp]
\centering
\includegraphics[scale=0.39]{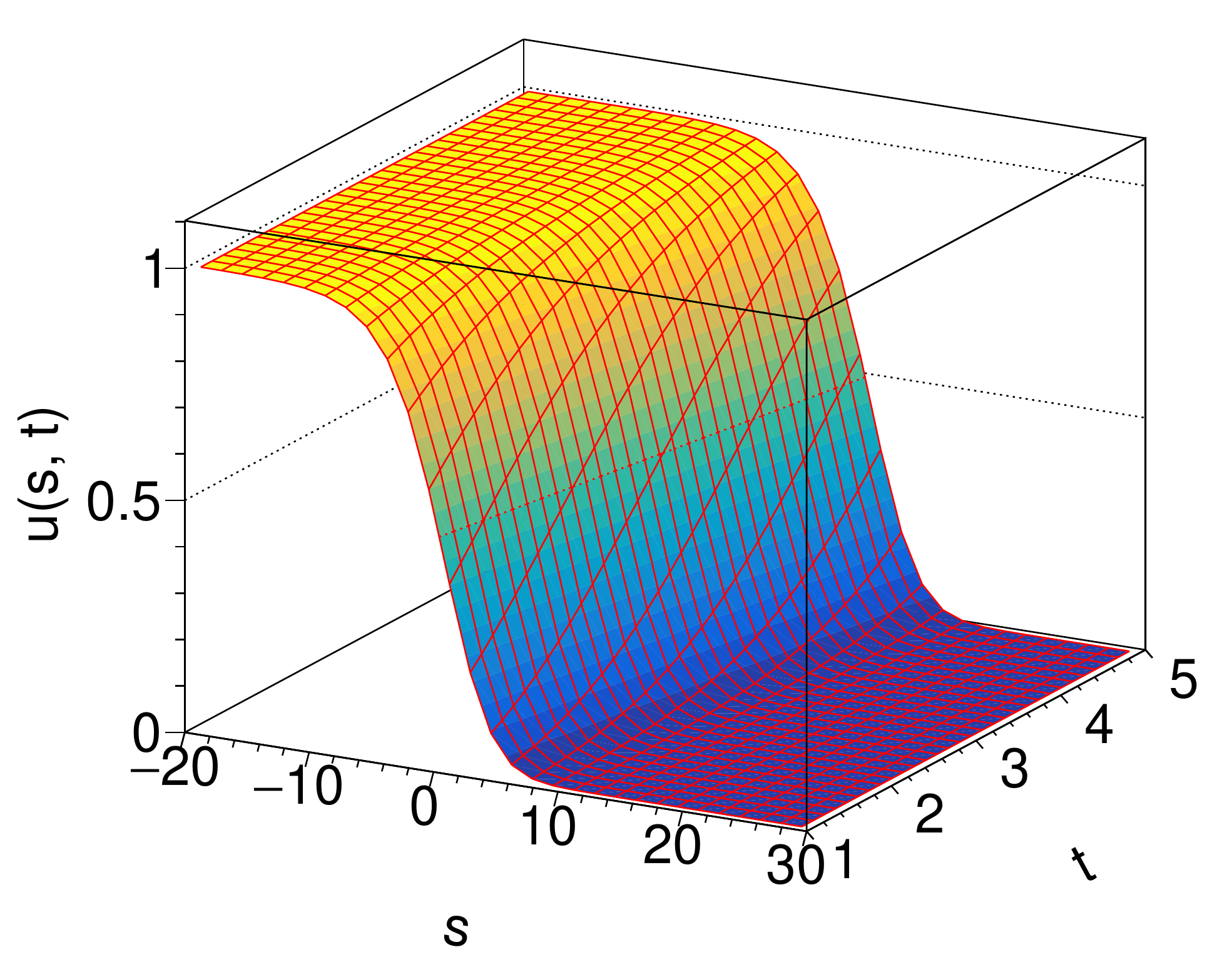}
\caption{The traveling wave solution of the FKPP equation from the homogeneous balance method,
with the coefficients $c_0 = 0$, $c_1 = 1/\sqrt{6}$, and $\theta = 0$. }
\label{fig:FKPP-traveling-wave}
\end{figure}

To solve the FKPP equation \cite{Yang:2020jmt}, we begin with a heuristic solution as,
\begin{equation}
\begin{split}
u(s,t) = \sum_{\rm m+n=1}^{\rm N} c_{\rm m+n} \frac{\partial^{\rm m+n}f(\omega(s,t))}{\partial^{\rm m}s\partial^{\rm n}t}.
\end{split}
\label{eq:HeuristicSolution}
\end{equation}
According to the partial balance principle, the power of $\partial\omega/\partial s$ should be balanced,
and the power of $\partial\omega/\partial t$ should also be balanced.
Hence, we obtain the following constraint,
\begin{equation}
\begin{split}
N = 2.
\end{split}
\label{eq:PowerConstraints}
\end{equation}
The heuristic solution now is written as,
\begin{equation}
\begin{split}
u(s,t) = f^{\prime\prime}(\omega)\left(\frac{\partial}{\partial s}\omega\right)^2
+ f^{\prime}(\omega)\frac{\partial^2}{\partial^2 s}\omega + c_1 f^{\prime}(\omega)\frac{\partial}{\partial s}\omega + c_0.
\end{split}
\label{eq:Solution-v1}
\end{equation}
Applying the above solution into the FKPP equation (Eq. (\ref{eq:FKPP})) again,
and apply the homogeneous balance principle, we get a differential equation of $f(\omega)$,
\begin{equation}
\begin{split}
(f^{\prime\prime}(\omega))^2 - f^{(4)}(\omega) = 0.
\end{split}
\label{eq:f-DifferentialEq}
\end{equation}
A particular solution for $f$ is then solved to be,
\begin{equation}
\begin{split}
f(\omega) = -6{\rm ln}\omega.
\end{split}
\label{eq:f-Solution}
\end{equation}
Using the FKPP equation and the homogeneous balance principle once again,
we get a differential equation of $\omega(s,t)$.
By solving the differential equation of $\omega$,
we get a traveling wave solution as a solitary wave,
\begin{equation}
\begin{split}
\omega(s,t)=1+e^{\kappa s + \beta t + \theta}.
\end{split}
\label{eq:omega-Solution}
\end{equation}
Inserting the solution into the FKPP equation,
we get the following constraints for the coefficients in the solution,
\begin{equation}
\begin{split}
c_0 = 0,1,\\
c_1 = \pm \frac{1}{\sqrt{6}},\\
\kappa = \frac{2c_0 - 1}{6c_1},\\
\beta = -5 c_1 \kappa,
\end{split}
\label{eq:CoefficientsConstraints}
\end{equation}
while $\theta$ is still a free parameter.
We choose $c_0=0$ and $c_1=1/\sqrt{6}$, in order to meet the physical result
that there is the strong absorption for the scattering amplitude at the very large rapidity.
Finally, we get an analytical solution for the FKPP equation, which is written as \cite{Yang:2020jmt},
\begin{equation}
\begin{split}
u(s,t)=\left[ \frac{e^{-s/\sqrt{6}+5t/6 + \theta }}{1 + e^{-s/\sqrt{6}+5t/6 + \theta }}  \right]^2.
\end{split}
\label{eq:FKPPSolutionFinal}
\end{equation}
Figure \ref{fig:FKPP-traveling-wave} shows the obtained analytical solution for the
FKPP equation as a function of $s$ and $t$.

\section{Definitive solution of BK equation by matching non-saturating gluon distribution}
\label{sec:BKSolution}

Before going onto the solution of BK equation,
let's have a look at the simplification of the BK equation again.
Fixing the running strong coupling constant and expanding the BFKL kernel
$\chi(\lambda)$ at $\lambda = \lambda_0$ \cite{Marquet:2005ic}, we get,
\begin{equation}
\begin{split}
A_0{\cal N} -{\cal N}^2 -\frac{\partial {\cal N}}{\partial Y} -A_1 \frac{\partial {\cal N}}{\partial L}
+\sum_{p=2}^{P} (-1)^p A_p \frac{\partial^p {\cal N}}{\partial^p L} = 0,
\end{split}
\label{eq:BK-expansion}
\end{equation}
with the coefficients derived as \cite{Marquet:2005ic},
\begin{equation}
\begin{split}
A_{p} = \sum_{i=0}^{P-p} (-1)^i\frac{\chi^{(i+p)}(\lambda_0)}{i!p!} \lambda_0^i.
\end{split}
\label{eq:BK-expansion-paras}
\end{equation}
In the so-called diffusive approximation (keeping the first three terms of the expansion),
the BK equation turns into the FKPP equation as suggested by Munier and Peschanski.
With $P=2$, we get the simplified BK equation \cite{Marquet:2005ic},
\begin{equation}
\begin{split}
A_0{\cal N} -{\cal N}^2 -\frac{\partial {\cal N}}{\partial Y} -A_1 \frac{\partial {\cal N}}{\partial L}
+ A_2 \frac{\partial^2 {\cal N}}{\partial^2 L} = 0.
\end{split}
\label{eq:BK-expansion2}
\end{equation}
Note that some variations of the values of $A_p$ (Eq. (\ref{eq:BK-expansion-paras})) are allowed, as the expansion point $\lambda_0$ is arbitrary,
the strong coupling $\alpha_{\rm s}(k^2)$ is slightly running,
and the truncation of the diffusive approximation may introduce some corrections.

Based on the analytical solution of FKPP equation discussed in the above section,
we obtain an analytical solution of the BK equation through
some variable transformations, which is written as,
\begin{equation}
\begin{split}
{\cal N}(L, Y) = \frac{ A_0 e^{\frac{5A_0Y}{3}}}{\left[e^{\frac{5A_0Y}{6}} +e^{[-\theta+\sqrt{\frac{A_0}{6A_2}}(L-A_1 Y)]} \right]^2}.
\end{split}
\label{eq:BK-solution}
\end{equation}
The geometric scaling of the virtual photon-proton cross section
is expressed as $\sigma^{\rm \gamma^*p}(Y,Q^2) = \sigma^{\rm \gamma^*p}(\tau)$,
which is a function of only one dimensionless variable $\tau=Q^2R_0^2$ in $x<0.01$ region,
where $R_0$ is the saturation radius, and $R_0=1/Q_{\rm s}$.
In terms of the forward scattering amplitude, the scaling property at low $k$ takes the form,
\begin{equation}
\begin{split}
{\cal N}(L={\rm ln}(k^2/k_0^2), Y) = {\cal N}\left(\frac{k^2}{Q_{\rm s}^2(Y)} \right),
\end{split}
\label{eq:BK-solution-scaling}
\end{equation}
where unit of Y is $\bar{\alpha}$.
Rewriting the solution ${\cal N}(L,Y)$ as ${\cal N}(k^2/Q_{\rm s}^2)$,
\begin{equation}
{\cal N}\left(\frac{k^2}{Q_{\rm s}^2(Y)}\right) = \frac{A_0}{\left[1+e^{-\theta}\left(\frac{k^2}{k_0^2e^{\left(A_1+5\sqrt{\frac{A_2A_0}{6}}\right)Y}}\right)^{\sqrt{\frac{A_0}{6A_2}}}\right]^2},
\label{eq:BK-solution-re}
\end{equation}
 then we extract the saturation scale to be,
\begin{equation}
\begin{split}
Q_{\rm s}^2(Y) = k_0^2 e^{\left(A_1+5\sqrt{A_0A_2/6} \right)Y}.
\end{split}
\label{eq:BK-saturation-scale}
\end{equation}
In the following analysis, we take $k_0^2$ to be $\Lambda_{\rm QCD}^2=0.04$ GeV$^2$ as
the reference point.

\begin{figure}[htbp]
\centering
\includegraphics[scale=0.38]{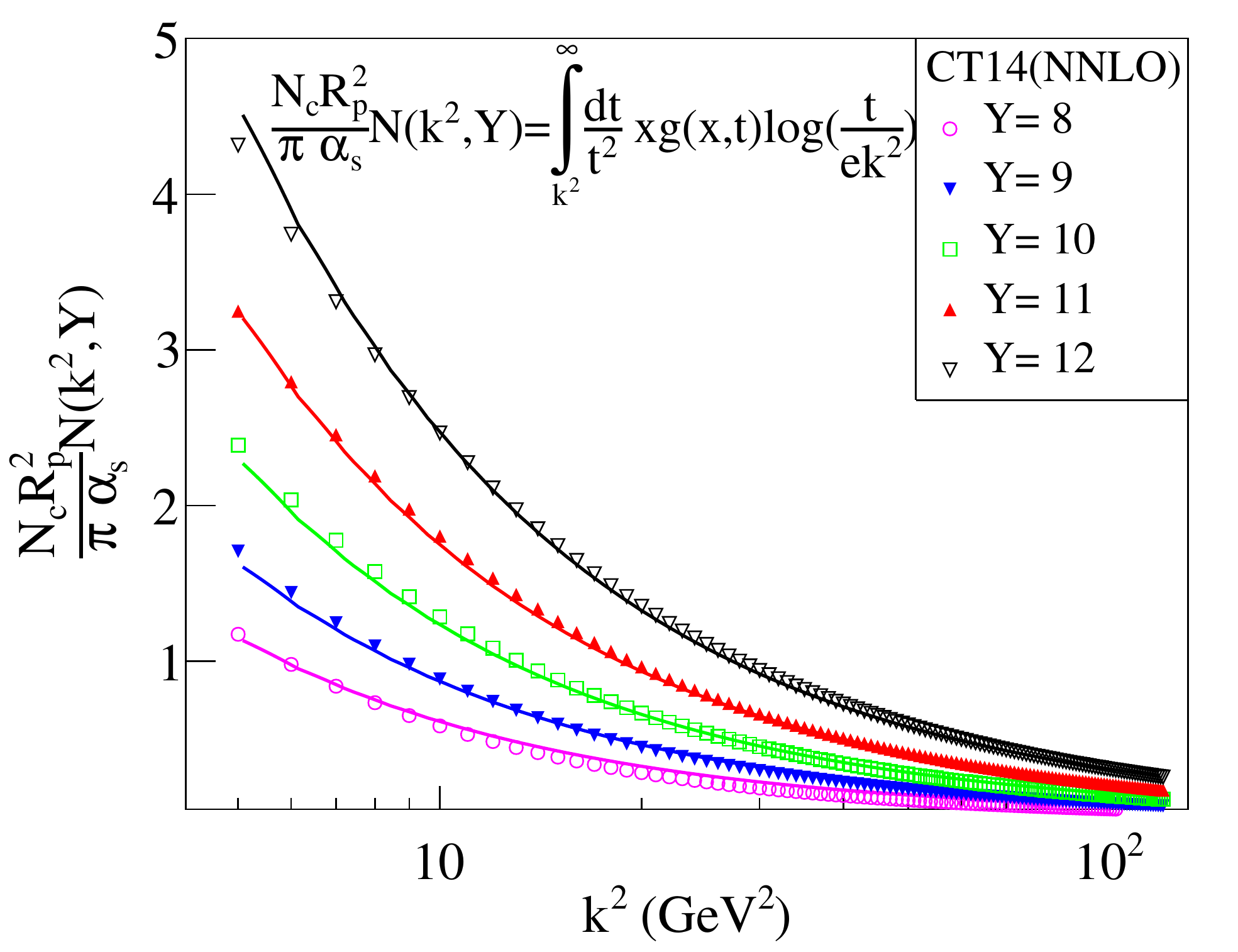}
\caption{The BK forward scattering amplitude is fitted by the gluon distributions of CT14(NNLO) at high $k^2$.
The curves show the fitted BK solutions, and the various markers show the calculations from CT14(NNLO) gluon distributions. }
\label{fig:amplitude-CT14-fit}
\end{figure}

In order to get the definitive solution of the BK equation for proton structure at small $x$,
the coefficients $A_0$, $A_1$ and $A_2$ should be determined.
By matching the dipole-proton cross section to
the normal DIS cross section in the parton model,
the dipole scattering amplitude is connected to the gluon distribution via \cite{Marquet:2005ic},
\begin{equation}
\begin{split}
{\cal N}(k, Y) = \frac{4\pi\alpha_{\rm s}}{N_{\rm c}R_{\rm p}^2} \int_{k}^{\infty} \frac{dp}{p}
\left[\frac{\partial}{\partial p^2}xg(x,p^2)\right] {\rm ln}\left(\frac{p}{k} \right)\\
= \frac{\pi\alpha_{\rm s}}{N_{\rm c}R_{\rm p}^2} \int_{k^2}^{\infty} \frac{dt}{t^2}
xg(x,t) {\rm ln}\left(\frac{t}{ek^2} \right).
\end{split}
\label{eq:amplitude-and-gluon}
\end{equation}
With the above formula, we can fix the values of $A_p$, by performing the fits
between the dipole scattering amplitude and the widely used global analyses
of the proton gluon distributions in the high $k^2$ ($>5$ GeV$^2$) region.
The relation in Eq. (\ref{eq:amplitude-and-gluon}) is based on an approximation that
the derivative of gluon distribution contains the information of the un-integrated
gluon distribution function.
And by a fit to the gluon distributions of CT14(NNLO),
we get the coefficients for the BK solution to be $A_0=33.3$, $A_1=-58.3$, $A_2=26.2$ and $\theta=-3.09$.
The fit to the CT14 parton distribution functions (PDFs) is shown in Fig. \ref{fig:amplitude-CT14-fit},
which displays a good fitting quality.

\begin{figure}[htbp]
\centering
\includegraphics[scale=0.38]{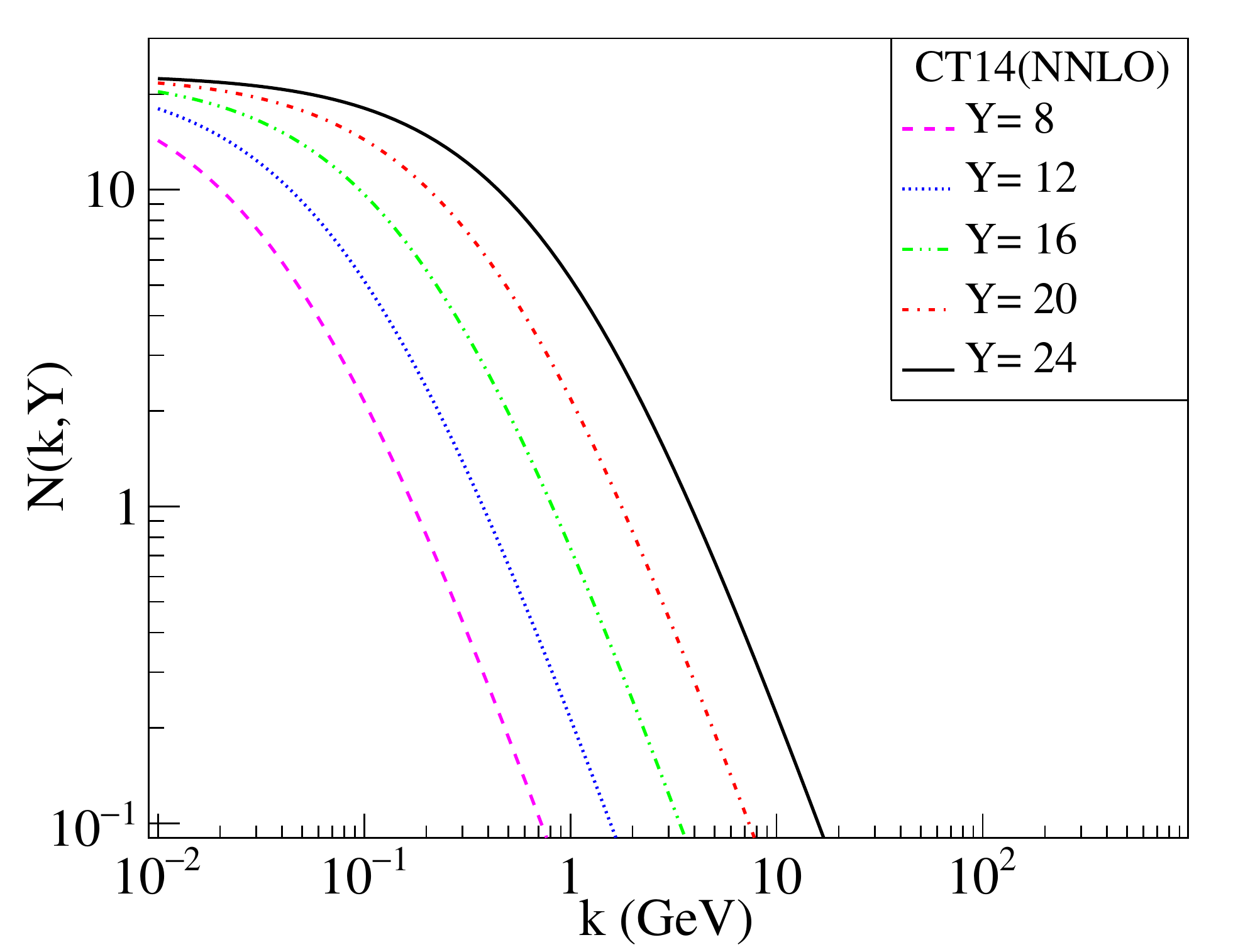}
\caption{The BK forward amplitude in the wide $k$ range at different rapidities,
with the parameters $A_0=33.3$, $A_1=-58.3$, $A_2=26.2$ and $\theta=-3.09$
by matching to CT14 PDFs in the non-saturation region. }
\label{fig:amplitude-CT14}
\end{figure}

\begin{figure}[htbp]
\centering
\includegraphics[scale=0.38]{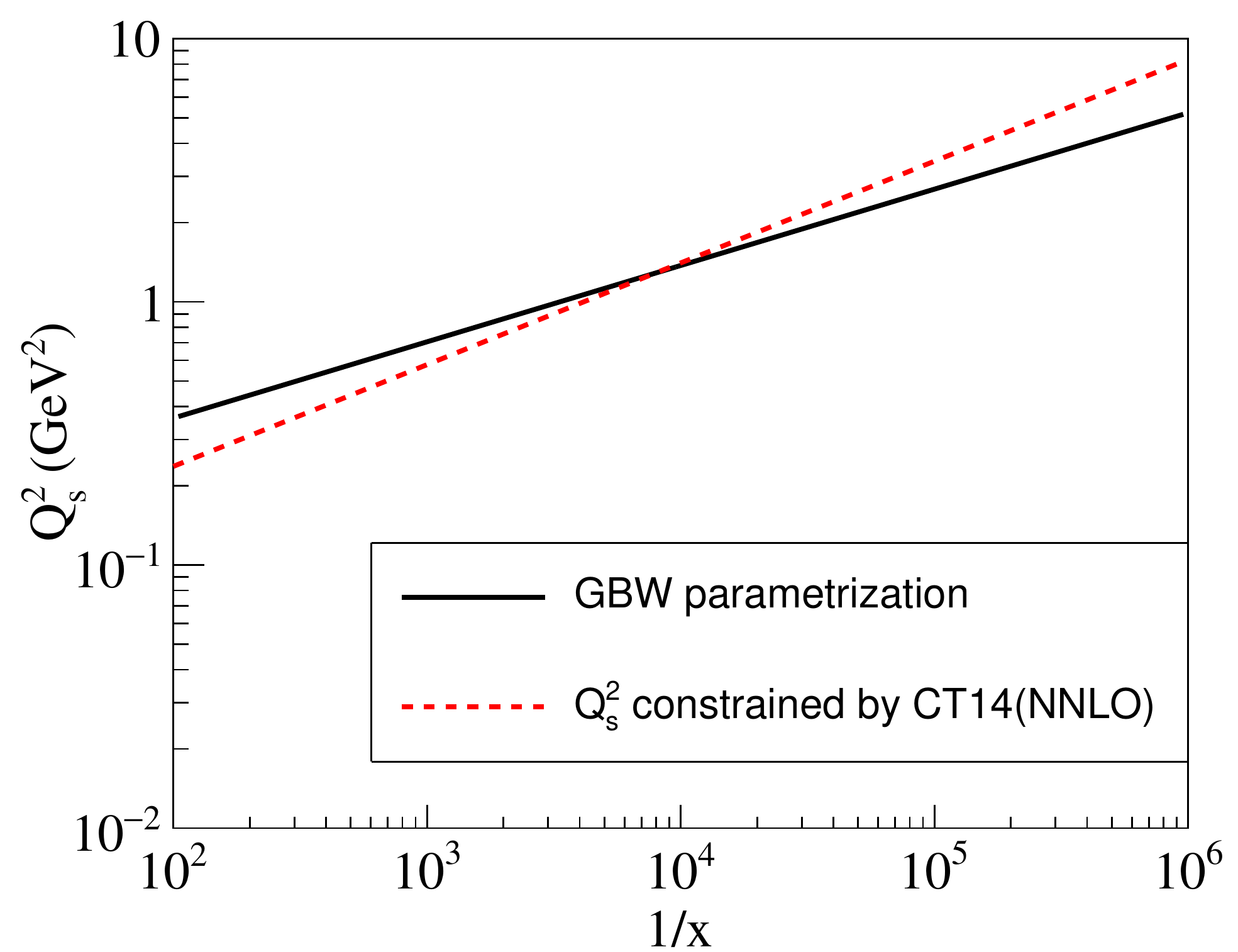}
\caption{The saturation scales extracted from the solution of BK equation in this work (dashed line),
compared to the prediction by GBW model (solid line) \cite{GolecBiernat:1998js,Bartels:2002cj,Kowalski:2003hm}. }
\label{fig:saturation-scales}
\end{figure}

\begin{figure}[htbp]
	\centering
	\includegraphics[scale=0.39]{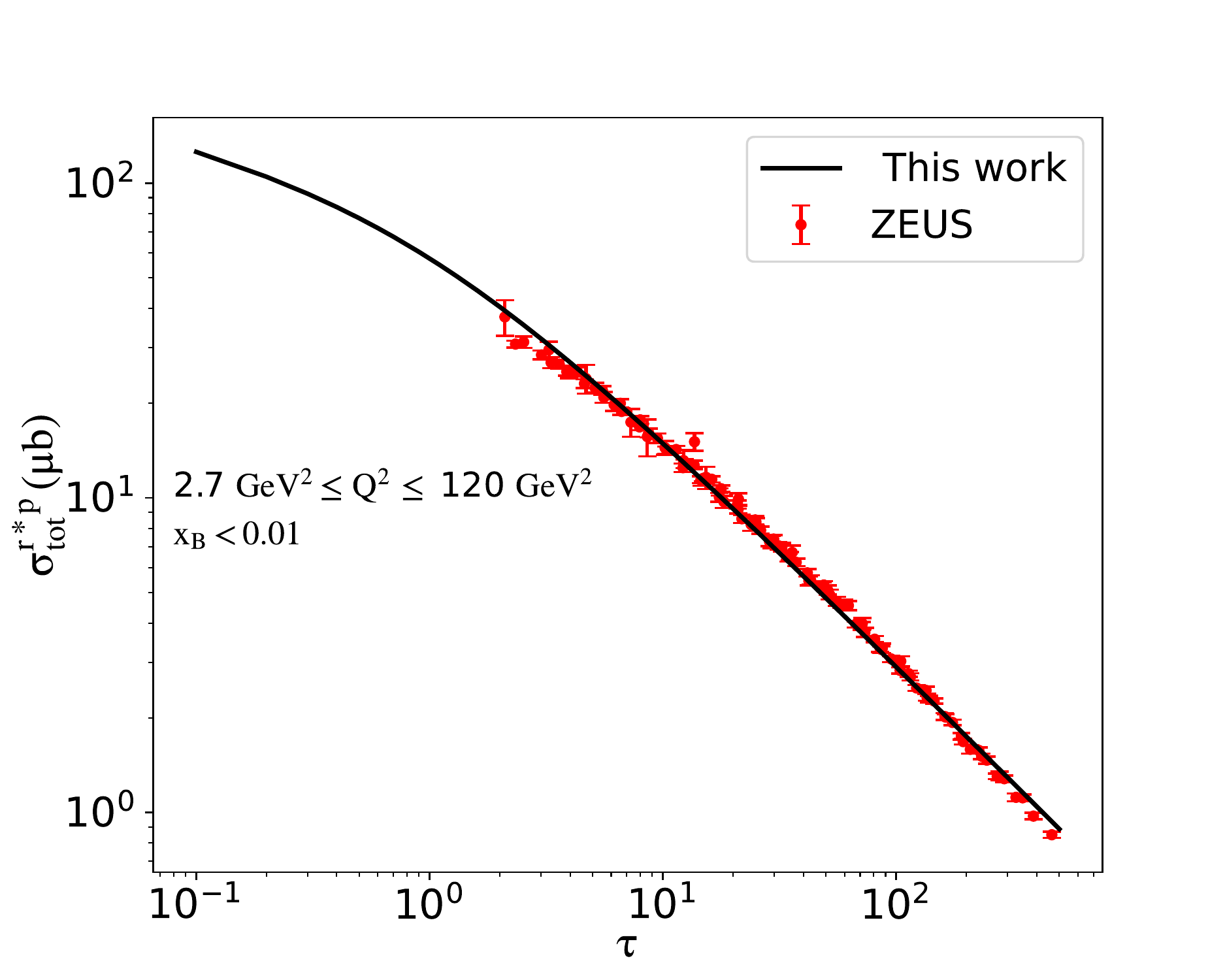}
	\caption{Comparison between our fit with the obtained saturation scale $Q_s$ (solid curve)
and the experimental data of ZEUS \cite{Breitweg:2000yn,Chekanov:2001qu} (red dots)
for the $\gamma^*p$ DIS cross section as a function of $\tau$.}
	\label{fig:geometric-scaling}
\end{figure}

With the definitive solution of the BK equation determined above,
we present the dipole-proton forward scattering amplitude in the full momentum range.
Fig. \ref{fig:amplitude-CT14} shows the BK scattering
amplitude at different rapidities constrained by CT14 PDFs.
It is very clear to see that these solutions exhibit the saturation behavior at the large rapidities.
To demonstrate where the saturation region is, the saturation scale $Q_{\rm s}^2(Y={\rm ln}(1/x))$
is usually used, which is rapidity-dependent. The saturation scale $Q_{\rm s}^2$ is viewed as
the boundary connecting the saturation domain and the non-saturation domain of the partons.
Fig. \ref{fig:saturation-scales} shows the saturation scales extracted by the BK solution determined
in this work. Our prediction indicates a very similar domain of the parton saturation
to the GBW parametrization. Our prediction gives $Q_{\rm s}^2(x)\propto (1/x)^{0.386}$,
and GBW prediction gives $Q_{\rm s}^2(x)\propto (1/x)^{0.29}$.
The existence of the parton saturation inside proton is understandable in the framework of
nonlinear evolution equation, and it is a well-established mechanism
for the geometric scaling phenomenon observed in experiment.

To test our obtained saturation scale $Q_{\rm s}^2(x)$,
we demonstrate the geometric scaling of the DIS cross section
with our solution.
The total $\gamma^* p$ cross section can be written as \cite{Stasto:2000er},
\begin{equation}
	\sigma_{T, L}\left(x, Q^{2}\right)=\int d^{2} \mathbf{r} \int_{0}^{1} d z\left|\Psi_{T, L}\left(r, z, Q^{2}\right)\right|^{2} \hat{\sigma}(r, x),
	\label{eq:sigma_gamma_p}
\end{equation}
where $\Psi_{T, L}$ is the wave function of
a longitudinal or a transverse virtual photon splitting into a dipole,
and $\hat{\sigma}$ is the dipole cross section.
Respectively, $\Psi_{T, L}$ and $\hat{\sigma}$ can be written as \cite{Stasto:2000er},
\begin{equation}
	\begin{aligned}
		\left|\Psi_{T}\right|^{2}=& \frac{3 \alpha_{e m}}{2 \pi^{2}} \sum_{f} e_{f}^{2}\left\{\left[z^{2}+(1-z)^{2}\right] \bar{Q}_{f}^{2} K_{1}^{2}\left(\bar{Q}_{f} r\right)\right.\\
		&\left.+m_{f}^{2} K_{0}^{2}\left(\bar{Q}_{f} r\right)\right\}, \\
		\left|\Psi_{L}\right|^{2}=& \frac{3 \alpha_{e m}}{2 \pi^{2}} \sum_{f} e_{f}^{2}\left\{4 Q^{2} z^{2}(1-z)^{2} K_{0}^{2}\left(\bar{Q}_{f} r\right)\right\},\\
	    \hat{\sigma}(r, x)=&\sigma_{0} g\left(\frac{r}{R_{0}(x)}\right),
	\end{aligned}
	\label{eq:wave_function_and_hat_sigma}
\end{equation}
in which $\bar{Q}_{f}^{2}=z(1-z) Q^{2}+m_{f}^{2}$, $K_{0}$ and $K_{1}$
are the modified Bessel functions and $m_f$ is the quark mass \cite{Stasto:2000er}.
$\sigma_{0}$ is a free parameter of normalization.
Then the geometric scaling \cite{Stasto:2000er,Armesto:2004ud,Ben:2017xny}
of the virtual photon-proton cross section is derived as,
\begin{equation}
	\sigma^{\gamma^{*} p}\left(x, Q^{2}\right)=\bar{\sigma}_{0}\left[\gamma_{E}+\Gamma(0, \xi)+\ln \xi\right],
	\label{eq:scaling_curve}
\end{equation}	
where $\xi=a/\tau^{b}$, $\tau=Q^2/Q_s^2$, $\gamma_{E}$ is Euler constant,
and $\Gamma(0,\xi)$ is the incomplete $\Gamma$ function.
By matching the ZEUS data \cite{Breitweg:2000yn,Chekanov:2001qu} in small $x_B$ ($<0.01$)
and high $Q^2$ region ($2.7 \,\rm GeV^2 \leq Q^2 \leq 120 \, GeV^2$),
we get the parameters $a=2.246$, and $b=0.745$ with $\bar{\sigma}_{0}$ fixed
at 40.56 $\rm \mu b$ \cite{Armesto:2004ud,Ben:2017xny}.
The comparison of our prediction with the ZEUS data is shown in Fig. \ref{fig:geometric-scaling}.

\section{Discussions and summary}
\label{sec:summary}

We have shown the analytical solution of nonlinear BK equation,
with the application of a mathematical method of homogeneous balance principle.
The homogeneous balance method would have some broad applications in solving
the nonlinear evolution equations in high energy QCD.
By fitting to the gluon distributions in the non-saturation region (high $Q^2$),
we have obtained the physical BK solution for the proton.
The solution in this work is similar to the traveling wave front introduced by Munier and Peschanski \cite{Munier:2003vc,Munier:2003sj,Munier:2004xu}.
The saturation is clearly shown in the scattering amplitude solution in the momentum space.
The saturation scales $Q_{\rm s}^2(Y)$ are also provided from our analysis.
This solution with the saturation behavior is an important dynamical mechanism
to explain the observed geometric scaling of the proton structure function at small $x$.
The BK equation with the truncation of BFKL kernel is successful
in interpreting the deep inelastic scattering data at current and closed accelerator facilities.

The solution of BK equation determined in this work is useful for the phenomenological
studies on the unitarization of high energy cross section, the parton saturation, and the small-$x$ physics.
Our results may be applied in the deep inelastic scattering \cite{Golec-Biernat:2017lfv},
the diffractive process \cite{Enberg:2004rz,Mueller:2018zwx,Cai:2020exu},
and the hadron collisions at high energy \cite{JalilianMarian:2005jf},
where the high parton density effect should be considered \cite{Gelis:2010nm}.
In US, the future Electron-Ion Collider (EIC) \cite{Accardi:2012qut} will be an important machine to study the phenomenology
of parton saturation physics. The saturation scale predictions can be evaluated with the future experiments.
For the proposed low energy Electron-ion collider in China (EicC) \cite{Chen:2018wyz,Chen:2020ijn},
the DIS cross section measurement
at low $Q^2$ and small $x_{\rm B}$ is also helpful for us to understand the saturation phenomenon
of partons inside proton or nucleus. The future high energy accelerator facilities are needed
for a precise understanding of the complex dynamics of strong interaction
and the high energy QCD frontiers.

\begin{acknowledgments}
We thank Profs. Kong-Qing YANG, Wei ZHU, Fan WANG and Yaping XIE for the suggestions and the fruitful discussions.
This work is supported by the Strategic Priority Research Program of Chinese Academy of Sciences
under the Grant NO. XDB34030301.
\end{acknowledgments}

\bibliographystyle{apsrev4-1}
\bibliography{refs}

\begin{thebibliography}{55}%
\makeatletter
\providecommand \@ifxundefined [1]{%
 \@ifx{#1\undefined}
}%
\providecommand \@ifnum [1]{%
 \ifnum #1\expandafter \@firstoftwo
 \else \expandafter \@secondoftwo
 \fi
}%
\providecommand \@ifx [1]{%
 \ifx #1\expandafter \@firstoftwo
 \else \expandafter \@secondoftwo
 \fi
}%
\providecommand \natexlab [1]{#1}%
\providecommand \enquote  [1]{``#1''}%
\providecommand \bibnamefont  [1]{#1}%
\providecommand \bibfnamefont [1]{#1}%
\providecommand \citenamefont [1]{#1}%
\providecommand \href@noop [0]{\@secondoftwo}%
\providecommand \href [0]{\begingroup \@sanitize@url \@href}%
\providecommand \@href[1]{\@@startlink{#1}\@@href}%
\providecommand \@@href[1]{\endgroup#1\@@endlink}%
\providecommand \@sanitize@url [0]{\catcode `\\12\catcode `\$12\catcode
  `\&12\catcode `\#12\catcode `\^12\catcode `\_12\catcode `\%12\relax}%
\providecommand \@@startlink[1]{}%
\providecommand \@@endlink[0]{}%
\providecommand \url  [0]{\begingroup\@sanitize@url \@url }%
\providecommand \@url [1]{\endgroup\@href {#1}{\urlprefix }}%
\providecommand \urlprefix  [0]{URL }%
\providecommand \Eprint [0]{\href }%
\providecommand \doibase [0]{http://dx.doi.org/}%
\providecommand \selectlanguage [0]{\@gobble}%
\providecommand \bibinfo  [0]{\@secondoftwo}%
\providecommand \bibfield  [0]{\@secondoftwo}%
\providecommand \translation [1]{[#1]}%
\providecommand \BibitemOpen [0]{}%
\providecommand \bibitemStop [0]{}%
\providecommand \bibitemNoStop [0]{.\EOS\space}%
\providecommand \EOS [0]{\spacefactor3000\relax}%
\providecommand \BibitemShut  [1]{\csname bibitem#1\endcsname}%
\let\auto@bib@innerbib\@empty
\bibitem [{\citenamefont {Lipatov}(1976)}]{Lipatov:1976zz}%
  \BibitemOpen
  \bibfield  {author} {\bibinfo {author} {\bibfnamefont {L.}~\bibnamefont
  {Lipatov}},\ }\href@noop {} {\bibfield  {journal} {\bibinfo  {journal} {Sov.
  J. Nucl. Phys.}\ }\textbf {\bibinfo {volume} {23}},\ \bibinfo {pages} {338}
  (\bibinfo {year} {1976})}\BibitemShut {NoStop}%
\bibitem [{\citenamefont {Kuraev}\ \emph {et~al.}(1977)\citenamefont {Kuraev},
  \citenamefont {Lipatov},\ and\ \citenamefont {Fadin}}]{Kuraev:1977fs}%
  \BibitemOpen
  \bibfield  {author} {\bibinfo {author} {\bibfnamefont {E.}~\bibnamefont
  {Kuraev}}, \bibinfo {author} {\bibfnamefont {L.}~\bibnamefont {Lipatov}}, \
  and\ \bibinfo {author} {\bibfnamefont {V.~S.}\ \bibnamefont {Fadin}},\
  }\href@noop {} {\bibfield  {journal} {\bibinfo  {journal} {Sov. Phys. JETP}\
  }\textbf {\bibinfo {volume} {45}},\ \bibinfo {pages} {199} (\bibinfo {year}
  {1977})}\BibitemShut {NoStop}%
\bibitem [{\citenamefont {Balitsky}\ and\ \citenamefont
  {Lipatov}(1978)}]{Balitsky:1978ic}%
  \BibitemOpen
  \bibfield  {author} {\bibinfo {author} {\bibfnamefont {I.}~\bibnamefont
  {Balitsky}}\ and\ \bibinfo {author} {\bibfnamefont {L.}~\bibnamefont
  {Lipatov}},\ }\href@noop {} {\bibfield  {journal} {\bibinfo  {journal} {Sov.
  J. Nucl. Phys.}\ }\textbf {\bibinfo {volume} {28}},\ \bibinfo {pages} {822}
  (\bibinfo {year} {1978})}\BibitemShut {NoStop}%
\bibitem [{\citenamefont {Froissart}(1961)}]{Froissart:1961ux}%
  \BibitemOpen
  \bibfield  {author} {\bibinfo {author} {\bibfnamefont {M.}~\bibnamefont
  {Froissart}},\ }\href {\doibase 10.1103/PhysRev.123.1053} {\bibfield
  {journal} {\bibinfo  {journal} {Phys. Rev.}\ }\textbf {\bibinfo {volume}
  {123}},\ \bibinfo {pages} {1053} (\bibinfo {year} {1961})}\BibitemShut
  {NoStop}%
\bibitem [{\citenamefont {Martin}(1963)}]{Martin:1962rt}%
  \BibitemOpen
  \bibfield  {author} {\bibinfo {author} {\bibfnamefont {A.}~\bibnamefont
  {Martin}},\ }\href {\doibase 10.1103/PhysRev.129.1432} {\bibfield  {journal}
  {\bibinfo  {journal} {Phys. Rev.}\ }\textbf {\bibinfo {volume} {129}},\
  \bibinfo {pages} {1432} (\bibinfo {year} {1963})}\BibitemShut {NoStop}%
\bibitem [{\citenamefont {Gribov}\ \emph {et~al.}(1981)\citenamefont {Gribov},
  \citenamefont {Levin},\ and\ \citenamefont {Ryskin}}]{Gribov:1981ac}%
  \BibitemOpen
  \bibfield  {author} {\bibinfo {author} {\bibfnamefont {L.}~\bibnamefont
  {Gribov}}, \bibinfo {author} {\bibfnamefont {E.}~\bibnamefont {Levin}}, \
  and\ \bibinfo {author} {\bibfnamefont {M.}~\bibnamefont {Ryskin}},\ }\href
  {\doibase 10.1016/0550-3213(81)90007-9} {\bibfield  {journal} {\bibinfo
  {journal} {Nucl. Phys. B}\ }\textbf {\bibinfo {volume} {188}},\ \bibinfo
  {pages} {555} (\bibinfo {year} {1981})}\BibitemShut {NoStop}%
\bibitem [{\citenamefont {Gribov}\ \emph {et~al.}(1983)\citenamefont {Gribov},
  \citenamefont {Levin},\ and\ \citenamefont {Ryskin}}]{Gribov:1984tu}%
  \BibitemOpen
  \bibfield  {author} {\bibinfo {author} {\bibfnamefont {L.}~\bibnamefont
  {Gribov}}, \bibinfo {author} {\bibfnamefont {E.}~\bibnamefont {Levin}}, \
  and\ \bibinfo {author} {\bibfnamefont {M.}~\bibnamefont {Ryskin}},\ }\href
  {\doibase 10.1016/0370-1573(83)90022-4} {\bibfield  {journal} {\bibinfo
  {journal} {Phys. Rept.}\ }\textbf {\bibinfo {volume} {100}},\ \bibinfo
  {pages} {1} (\bibinfo {year} {1983})}\BibitemShut {NoStop}%
\bibitem [{\citenamefont {Mueller}\ and\ \citenamefont
  {Qiu}(1986)}]{Mueller:1985wy}%
  \BibitemOpen
  \bibfield  {author} {\bibinfo {author} {\bibfnamefont {A.~H.}\ \bibnamefont
  {Mueller}}\ and\ \bibinfo {author} {\bibfnamefont {J.-w.}\ \bibnamefont
  {Qiu}},\ }\href {\doibase 10.1016/0550-3213(86)90164-1} {\bibfield  {journal}
  {\bibinfo  {journal} {Nucl. Phys. B}\ }\textbf {\bibinfo {volume} {268}},\
  \bibinfo {pages} {427} (\bibinfo {year} {1986})}\BibitemShut {NoStop}%
\bibitem [{\citenamefont {Mueller}(1990)}]{Mueller:1989st}%
  \BibitemOpen
  \bibfield  {author} {\bibinfo {author} {\bibfnamefont {A.~H.}\ \bibnamefont
  {Mueller}},\ }\href {\doibase 10.1016/0550-3213(90)90173-B} {\bibfield
  {journal} {\bibinfo  {journal} {Nucl. Phys. B}\ }\textbf {\bibinfo {volume}
  {335}},\ \bibinfo {pages} {115} (\bibinfo {year} {1990})}\BibitemShut
  {NoStop}%
\bibitem [{\citenamefont {Zhu}\ and\ \citenamefont {Ruan}(1999)}]{Zhu:1999ht}%
  \BibitemOpen
  \bibfield  {author} {\bibinfo {author} {\bibfnamefont {W.}~\bibnamefont
  {Zhu}}\ and\ \bibinfo {author} {\bibfnamefont {J.-h.}\ \bibnamefont {Ruan}},\
  }\href {\doibase 10.1016/S0550-3213(99)00461-7} {\bibfield  {journal}
  {\bibinfo  {journal} {Nucl. Phys. B}\ }\textbf {\bibinfo {volume} {559}},\
  \bibinfo {pages} {378} (\bibinfo {year} {1999})},\ \Eprint
  {http://arxiv.org/abs/hep-ph/9907330} {arXiv:hep-ph/9907330} \BibitemShut
  {NoStop}%
\bibitem [{\citenamefont {Balitsky}(1996)}]{Balitsky:1995ub}%
  \BibitemOpen
  \bibfield  {author} {\bibinfo {author} {\bibfnamefont {I.}~\bibnamefont
  {Balitsky}},\ }\href {\doibase 10.1016/0550-3213(95)00638-9} {\bibfield
  {journal} {\bibinfo  {journal} {Nucl. Phys. B}\ }\textbf {\bibinfo {volume}
  {463}},\ \bibinfo {pages} {99} (\bibinfo {year} {1996})},\ \Eprint
  {http://arxiv.org/abs/hep-ph/9509348} {arXiv:hep-ph/9509348} \BibitemShut
  {NoStop}%
\bibitem [{\citenamefont {Jalilian-Marian}\ \emph {et~al.}(1997)\citenamefont
  {Jalilian-Marian}, \citenamefont {Kovner}, \citenamefont {Leonidov},\ and\
  \citenamefont {Weigert}}]{JalilianMarian:1997jx}%
  \BibitemOpen
  \bibfield  {author} {\bibinfo {author} {\bibfnamefont {J.}~\bibnamefont
  {Jalilian-Marian}}, \bibinfo {author} {\bibfnamefont {A.}~\bibnamefont
  {Kovner}}, \bibinfo {author} {\bibfnamefont {A.}~\bibnamefont {Leonidov}}, \
  and\ \bibinfo {author} {\bibfnamefont {H.}~\bibnamefont {Weigert}},\ }\href
  {\doibase 10.1016/S0550-3213(97)00440-9} {\bibfield  {journal} {\bibinfo
  {journal} {Nucl. Phys. B}\ }\textbf {\bibinfo {volume} {504}},\ \bibinfo
  {pages} {415} (\bibinfo {year} {1997})},\ \Eprint
  {http://arxiv.org/abs/hep-ph/9701284} {arXiv:hep-ph/9701284} \BibitemShut
  {NoStop}%
\bibitem [{\citenamefont {Iancu}\ \emph
  {et~al.}(2001{\natexlab{a}})\citenamefont {Iancu}, \citenamefont {Leonidov},\
  and\ \citenamefont {McLerran}}]{Iancu:2000hn}%
  \BibitemOpen
  \bibfield  {author} {\bibinfo {author} {\bibfnamefont {E.}~\bibnamefont
  {Iancu}}, \bibinfo {author} {\bibfnamefont {A.}~\bibnamefont {Leonidov}}, \
  and\ \bibinfo {author} {\bibfnamefont {L.~D.}\ \bibnamefont {McLerran}},\
  }\href {\doibase 10.1016/S0375-9474(01)00642-X} {\bibfield  {journal}
  {\bibinfo  {journal} {Nucl. Phys. A}\ }\textbf {\bibinfo {volume} {692}},\
  \bibinfo {pages} {583} (\bibinfo {year} {2001}{\natexlab{a}})},\ \Eprint
  {http://arxiv.org/abs/hep-ph/0011241} {arXiv:hep-ph/0011241} \BibitemShut
  {NoStop}%
\bibitem [{\citenamefont {Weigert}(2002)}]{Weigert:2000gi}%
  \BibitemOpen
  \bibfield  {author} {\bibinfo {author} {\bibfnamefont {H.}~\bibnamefont
  {Weigert}},\ }\href {\doibase 10.1016/S0375-9474(01)01668-2} {\bibfield
  {journal} {\bibinfo  {journal} {Nucl. Phys. A}\ }\textbf {\bibinfo {volume}
  {703}},\ \bibinfo {pages} {823} (\bibinfo {year} {2002})},\ \Eprint
  {http://arxiv.org/abs/hep-ph/0004044} {arXiv:hep-ph/0004044} \BibitemShut
  {NoStop}%
\bibitem [{\citenamefont {Iancu}\ \emph
  {et~al.}(2001{\natexlab{b}})\citenamefont {Iancu}, \citenamefont {Leonidov},\
  and\ \citenamefont {McLerran}}]{Iancu:2001ad}%
  \BibitemOpen
  \bibfield  {author} {\bibinfo {author} {\bibfnamefont {E.}~\bibnamefont
  {Iancu}}, \bibinfo {author} {\bibfnamefont {A.}~\bibnamefont {Leonidov}}, \
  and\ \bibinfo {author} {\bibfnamefont {L.~D.}\ \bibnamefont {McLerran}},\
  }\href {\doibase 10.1016/S0370-2693(01)00524-X} {\bibfield  {journal}
  {\bibinfo  {journal} {Phys. Lett. B}\ }\textbf {\bibinfo {volume} {510}},\
  \bibinfo {pages} {133} (\bibinfo {year} {2001}{\natexlab{b}})},\ \Eprint
  {http://arxiv.org/abs/hep-ph/0102009} {arXiv:hep-ph/0102009} \BibitemShut
  {NoStop}%
\bibitem [{\citenamefont {Jalilian-Marian}\ and\ \citenamefont
  {Kovchegov}(2006)}]{JalilianMarian:2005jf}%
  \BibitemOpen
  \bibfield  {author} {\bibinfo {author} {\bibfnamefont {J.}~\bibnamefont
  {Jalilian-Marian}}\ and\ \bibinfo {author} {\bibfnamefont {Y.~V.}\
  \bibnamefont {Kovchegov}},\ }\href {\doibase 10.1016/j.ppnp.2005.07.002}
  {\bibfield  {journal} {\bibinfo  {journal} {Prog. Part. Nucl. Phys.}\
  }\textbf {\bibinfo {volume} {56}},\ \bibinfo {pages} {104} (\bibinfo {year}
  {2006})},\ \Eprint {http://arxiv.org/abs/hep-ph/0505052}
  {arXiv:hep-ph/0505052} \BibitemShut {NoStop}%
\bibitem [{\citenamefont {Gelis}\ \emph {et~al.}(2010)\citenamefont {Gelis},
  \citenamefont {Iancu}, \citenamefont {Jalilian-Marian},\ and\ \citenamefont
  {Venugopalan}}]{Gelis:2010nm}%
  \BibitemOpen
  \bibfield  {author} {\bibinfo {author} {\bibfnamefont {F.}~\bibnamefont
  {Gelis}}, \bibinfo {author} {\bibfnamefont {E.}~\bibnamefont {Iancu}},
  \bibinfo {author} {\bibfnamefont {J.}~\bibnamefont {Jalilian-Marian}}, \ and\
  \bibinfo {author} {\bibfnamefont {R.}~\bibnamefont {Venugopalan}},\ }\href
  {\doibase 10.1146/annurev.nucl.010909.083629} {\bibfield  {journal} {\bibinfo
   {journal} {Ann. Rev. Nucl. Part. Sci.}\ }\textbf {\bibinfo {volume} {60}},\
  \bibinfo {pages} {463} (\bibinfo {year} {2010})},\ \Eprint
  {http://arxiv.org/abs/1002.0333} {arXiv:1002.0333 [hep-ph]} \BibitemShut
  {NoStop}%
\bibitem [{\citenamefont {Balitsky}(1997)}]{Balitsky:1997mk}%
  \BibitemOpen
  \bibfield  {author} {\bibinfo {author} {\bibfnamefont {I.}~\bibnamefont
  {Balitsky}},\ }\href {\doibase 10.1063/1.53693} {\bibfield  {journal}
  {\bibinfo  {journal} {AIP Conf. Proc.}\ }\textbf {\bibinfo {volume} {407}},\
  \bibinfo {pages} {953} (\bibinfo {year} {1997})},\ \Eprint
  {http://arxiv.org/abs/hep-ph/9706411} {arXiv:hep-ph/9706411} \BibitemShut
  {NoStop}%
\bibitem [{\citenamefont {Kovchegov}(1999)}]{Kovchegov:1999yj}%
  \BibitemOpen
  \bibfield  {author} {\bibinfo {author} {\bibfnamefont {Y.~V.}\ \bibnamefont
  {Kovchegov}},\ }\href {\doibase 10.1103/PhysRevD.60.034008} {\bibfield
  {journal} {\bibinfo  {journal} {Phys. Rev. D}\ }\textbf {\bibinfo {volume}
  {60}},\ \bibinfo {pages} {034008} (\bibinfo {year} {1999})},\ \Eprint
  {http://arxiv.org/abs/hep-ph/9901281} {arXiv:hep-ph/9901281} \BibitemShut
  {NoStop}%
\bibitem [{\citenamefont {Kovchegov}(2000)}]{Kovchegov:1999ua}%
  \BibitemOpen
  \bibfield  {author} {\bibinfo {author} {\bibfnamefont {Y.~V.}\ \bibnamefont
  {Kovchegov}},\ }\href {\doibase 10.1103/PhysRevD.61.074018} {\bibfield
  {journal} {\bibinfo  {journal} {Phys. Rev. D}\ }\textbf {\bibinfo {volume}
  {61}},\ \bibinfo {pages} {074018} (\bibinfo {year} {2000})},\ \Eprint
  {http://arxiv.org/abs/hep-ph/9905214} {arXiv:hep-ph/9905214} \BibitemShut
  {NoStop}%
\bibitem [{\citenamefont {Balitsky}(2001)}]{Balitsky:2001re}%
  \BibitemOpen
  \bibfield  {author} {\bibinfo {author} {\bibfnamefont {I.}~\bibnamefont
  {Balitsky}},\ }\href {\doibase 10.1016/S0370-2693(01)01041-3} {\bibfield
  {journal} {\bibinfo  {journal} {Phys. Lett. B}\ }\textbf {\bibinfo {volume}
  {518}},\ \bibinfo {pages} {235} (\bibinfo {year} {2001})},\ \Eprint
  {http://arxiv.org/abs/hep-ph/0105334} {arXiv:hep-ph/0105334} \BibitemShut
  {NoStop}%
\bibitem [{\citenamefont {Armesto}\ and\ \citenamefont
  {Braun}(2001)}]{Armesto:2001fa}%
  \BibitemOpen
  \bibfield  {author} {\bibinfo {author} {\bibfnamefont {N.}~\bibnamefont
  {Armesto}}\ and\ \bibinfo {author} {\bibfnamefont {M.}~\bibnamefont
  {Braun}},\ }\href {\doibase 10.1007/s100520100685} {\bibfield  {journal}
  {\bibinfo  {journal} {Eur. Phys. J. C}\ }\textbf {\bibinfo {volume} {20}},\
  \bibinfo {pages} {517} (\bibinfo {year} {2001})},\ \Eprint
  {http://arxiv.org/abs/hep-ph/0104038} {arXiv:hep-ph/0104038} \BibitemShut
  {NoStop}%
\bibitem [{\citenamefont {Golec-Biernat}\ \emph {et~al.}(2002)\citenamefont
  {Golec-Biernat}, \citenamefont {Motyka},\ and\ \citenamefont
  {Stasto}}]{GolecBiernat:2001if}%
  \BibitemOpen
  \bibfield  {author} {\bibinfo {author} {\bibfnamefont {K.~J.}\ \bibnamefont
  {Golec-Biernat}}, \bibinfo {author} {\bibfnamefont {L.}~\bibnamefont
  {Motyka}}, \ and\ \bibinfo {author} {\bibfnamefont {A.}~\bibnamefont
  {Stasto}},\ }\href {\doibase 10.1103/PhysRevD.65.074037} {\bibfield
  {journal} {\bibinfo  {journal} {Phys. Rev. D}\ }\textbf {\bibinfo {volume}
  {65}},\ \bibinfo {pages} {074037} (\bibinfo {year} {2002})},\ \Eprint
  {http://arxiv.org/abs/hep-ph/0110325} {arXiv:hep-ph/0110325} \BibitemShut
  {NoStop}%
\bibitem [{\citenamefont {Enberg}\ and\ \citenamefont
  {Peschanski}(2006)}]{Enberg:2005zj}%
  \BibitemOpen
  \bibfield  {author} {\bibinfo {author} {\bibfnamefont {R.}~\bibnamefont
  {Enberg}}\ and\ \bibinfo {author} {\bibfnamefont {R.~B.}\ \bibnamefont
  {Peschanski}},\ }\href {\doibase 10.1016/j.nuclphysa.2005.12.012} {\bibfield
  {journal} {\bibinfo  {journal} {Nucl. Phys. A}\ }\textbf {\bibinfo {volume}
  {767}},\ \bibinfo {pages} {189} (\bibinfo {year} {2006})},\ \Eprint
  {http://arxiv.org/abs/hep-ph/0510352} {arXiv:hep-ph/0510352} \BibitemShut
  {NoStop}%
\bibitem [{\citenamefont {Marquet}\ and\ \citenamefont
  {Soyez}(2005)}]{Marquet:2005zf}%
  \BibitemOpen
  \bibfield  {author} {\bibinfo {author} {\bibfnamefont {C.}~\bibnamefont
  {Marquet}}\ and\ \bibinfo {author} {\bibfnamefont {G.}~\bibnamefont
  {Soyez}},\ }\href {\doibase 10.1016/j.nuclphysa.2005.05.198} {\bibfield
  {journal} {\bibinfo  {journal} {Nucl. Phys. A}\ }\textbf {\bibinfo {volume}
  {760}},\ \bibinfo {pages} {208} (\bibinfo {year} {2005})},\ \Eprint
  {http://arxiv.org/abs/hep-ph/0504080} {arXiv:hep-ph/0504080} \BibitemShut
  {NoStop}%
\bibitem [{\citenamefont {Munier}\ and\ \citenamefont
  {Peschanski}(2003)}]{Munier:2003vc}%
  \BibitemOpen
  \bibfield  {author} {\bibinfo {author} {\bibfnamefont {S.}~\bibnamefont
  {Munier}}\ and\ \bibinfo {author} {\bibfnamefont {R.~B.}\ \bibnamefont
  {Peschanski}},\ }\href {\doibase 10.1103/PhysRevLett.91.232001} {\bibfield
  {journal} {\bibinfo  {journal} {Phys. Rev. Lett.}\ }\textbf {\bibinfo
  {volume} {91}},\ \bibinfo {pages} {232001} (\bibinfo {year} {2003})},\
  \Eprint {http://arxiv.org/abs/hep-ph/0309177} {arXiv:hep-ph/0309177}
  \BibitemShut {NoStop}%
\bibitem [{\citenamefont {Munier}\ and\ \citenamefont
  {Peschanski}(2004{\natexlab{a}})}]{Munier:2003sj}%
  \BibitemOpen
  \bibfield  {author} {\bibinfo {author} {\bibfnamefont {S.}~\bibnamefont
  {Munier}}\ and\ \bibinfo {author} {\bibfnamefont {R.~B.}\ \bibnamefont
  {Peschanski}},\ }\href {\doibase 10.1103/PhysRevD.69.034008} {\bibfield
  {journal} {\bibinfo  {journal} {Phys. Rev. D}\ }\textbf {\bibinfo {volume}
  {69}},\ \bibinfo {pages} {034008} (\bibinfo {year} {2004}{\natexlab{a}})},\
  \Eprint {http://arxiv.org/abs/hep-ph/0310357} {arXiv:hep-ph/0310357}
  \BibitemShut {NoStop}%
\bibitem [{\citenamefont {Munier}\ and\ \citenamefont
  {Peschanski}(2004{\natexlab{b}})}]{Munier:2004xu}%
  \BibitemOpen
  \bibfield  {author} {\bibinfo {author} {\bibfnamefont {S.}~\bibnamefont
  {Munier}}\ and\ \bibinfo {author} {\bibfnamefont {R.~B.}\ \bibnamefont
  {Peschanski}},\ }\href {\doibase 10.1103/PhysRevD.70.077503} {\bibfield
  {journal} {\bibinfo  {journal} {Phys. Rev. D}\ }\textbf {\bibinfo {volume}
  {70}},\ \bibinfo {pages} {077503} (\bibinfo {year} {2004}{\natexlab{b}})},\
  \Eprint {http://arxiv.org/abs/hep-ph/0401215} {arXiv:hep-ph/0401215}
  \BibitemShut {NoStop}%
\bibitem [{\citenamefont {Xiang}\ \emph {et~al.}(2017)\citenamefont {Xiang},
  \citenamefont {Cai},\ and\ \citenamefont {Zhou}}]{Xiang:2017fjr}%
  \BibitemOpen
  \bibfield  {author} {\bibinfo {author} {\bibfnamefont {W.}~\bibnamefont
  {Xiang}}, \bibinfo {author} {\bibfnamefont {S.}~\bibnamefont {Cai}}, \ and\
  \bibinfo {author} {\bibfnamefont {D.}~\bibnamefont {Zhou}},\ }\href {\doibase
  10.1103/PhysRevD.95.116009} {\bibfield  {journal} {\bibinfo  {journal} {Phys.
  Rev. D}\ }\textbf {\bibinfo {volume} {95}},\ \bibinfo {pages} {116009}
  (\bibinfo {year} {2017})},\ \Eprint {http://arxiv.org/abs/1701.07378}
  {arXiv:1701.07378 [hep-ph]} \BibitemShut {NoStop}%
\bibitem [{\citenamefont {Xiang}\ \emph {et~al.}(2020)\citenamefont {Xiang},
  \citenamefont {Cai}, \citenamefont {Wang},\ and\ \citenamefont
  {Zhou}}]{Xiang:2019kre}%
  \BibitemOpen
  \bibfield  {author} {\bibinfo {author} {\bibfnamefont {W.}~\bibnamefont
  {Xiang}}, \bibinfo {author} {\bibfnamefont {Y.}~\bibnamefont {Cai}}, \bibinfo
  {author} {\bibfnamefont {M.}~\bibnamefont {Wang}}, \ and\ \bibinfo {author}
  {\bibfnamefont {D.}~\bibnamefont {Zhou}},\ }\href {\doibase
  10.1103/PhysRevD.101.076005} {\bibfield  {journal} {\bibinfo  {journal}
  {Phys. Rev. D}\ }\textbf {\bibinfo {volume} {101}},\ \bibinfo {pages}
  {076005} (\bibinfo {year} {2020})},\ \Eprint
  {http://arxiv.org/abs/1911.06744} {arXiv:1911.06744 [hep-ph]} \BibitemShut
  {NoStop}%
\bibitem [{\citenamefont {Iancu}\ \emph {et~al.}(2005)\citenamefont {Iancu},
  \citenamefont {Mueller},\ and\ \citenamefont {Munier}}]{Iancu:2004es}%
  \BibitemOpen
  \bibfield  {author} {\bibinfo {author} {\bibfnamefont {E.}~\bibnamefont
  {Iancu}}, \bibinfo {author} {\bibfnamefont {A.}~\bibnamefont {Mueller}}, \
  and\ \bibinfo {author} {\bibfnamefont {S.}~\bibnamefont {Munier}},\ }\href
  {\doibase 10.1016/j.physletb.2004.12.009} {\bibfield  {journal} {\bibinfo
  {journal} {Phys. Lett. B}\ }\textbf {\bibinfo {volume} {606}},\ \bibinfo
  {pages} {342} (\bibinfo {year} {2005})},\ \Eprint
  {http://arxiv.org/abs/hep-ph/0410018} {arXiv:hep-ph/0410018} \BibitemShut
  {NoStop}%
\bibitem [{\citenamefont {Munier}(2015)}]{Munier:2014bba}%
  \BibitemOpen
  \bibfield  {author} {\bibinfo {author} {\bibfnamefont {S.}~\bibnamefont
  {Munier}},\ }\href {\doibase 10.1007/s11433-015-5666-7} {\bibfield  {journal}
  {\bibinfo  {journal} {Sci. China Phys. Mech. Astron.}\ }\textbf {\bibinfo
  {volume} {58}},\ \bibinfo {pages} {81001} (\bibinfo {year} {2015})},\ \Eprint
  {http://arxiv.org/abs/1410.6478} {arXiv:1410.6478 [hep-ph]} \BibitemShut
  {NoStop}%
\bibitem [{\citenamefont {Mueller}\ and\ \citenamefont
  {Munier}(2018)}]{Mueller:2018zwx}%
  \BibitemOpen
  \bibfield  {author} {\bibinfo {author} {\bibfnamefont {A.}~\bibnamefont
  {Mueller}}\ and\ \bibinfo {author} {\bibfnamefont {S.}~\bibnamefont
  {Munier}},\ }\href {\doibase 10.1103/PhysRevLett.121.082001} {\bibfield
  {journal} {\bibinfo  {journal} {Phys. Rev. Lett.}\ }\textbf {\bibinfo
  {volume} {121}},\ \bibinfo {pages} {082001} (\bibinfo {year} {2018})},\
  \Eprint {http://arxiv.org/abs/1805.09417} {arXiv:1805.09417 [hep-ph]}
  \BibitemShut {NoStop}%
\bibitem [{\citenamefont {Enberg}(2005)}]{Enberg:2005uv}%
  \BibitemOpen
  \bibfield  {author} {\bibinfo {author} {\bibfnamefont {R.}~\bibnamefont
  {Enberg}},\ }\href {\doibase 10.1063/1.2122043} {\bibfield  {journal}
  {\bibinfo  {journal} {AIP Conf. Proc.}\ }\textbf {\bibinfo {volume} {792}},\
  \bibinfo {pages} {307} (\bibinfo {year} {2005})},\ \Eprint
  {http://arxiv.org/abs/hep-ph/0507153} {arXiv:hep-ph/0507153} \BibitemShut
  {NoStop}%
\bibitem [{\citenamefont {Stasto}\ \emph {et~al.}(2001)\citenamefont {Stasto},
  \citenamefont {Golec-Biernat},\ and\ \citenamefont
  {Kwiecinski}}]{Stasto:2000er}%
  \BibitemOpen
  \bibfield  {author} {\bibinfo {author} {\bibfnamefont {A.}~\bibnamefont
  {Stasto}}, \bibinfo {author} {\bibfnamefont {K.~J.}\ \bibnamefont
  {Golec-Biernat}}, \ and\ \bibinfo {author} {\bibfnamefont {J.}~\bibnamefont
  {Kwiecinski}},\ }\href {\doibase 10.1103/PhysRevLett.86.596} {\bibfield
  {journal} {\bibinfo  {journal} {Phys. Rev. Lett.}\ }\textbf {\bibinfo
  {volume} {86}},\ \bibinfo {pages} {596} (\bibinfo {year} {2001})},\ \Eprint
  {http://arxiv.org/abs/hep-ph/0007192} {arXiv:hep-ph/0007192} \BibitemShut
  {NoStop}%
\bibitem [{\citenamefont {de~Santana~Amaral}\ \emph
  {et~al.}(2007{\natexlab{a}})\citenamefont {de~Santana~Amaral}, \citenamefont
  {Betemps}, \citenamefont {Gay~Ducati},\ and\ \citenamefont
  {Soyez}}]{deSantanaAmaral:2007zzb}%
  \BibitemOpen
  \bibfield  {author} {\bibinfo {author} {\bibfnamefont {J.}~\bibnamefont
  {de~Santana~Amaral}}, \bibinfo {author} {\bibfnamefont {M.}~\bibnamefont
  {Betemps}}, \bibinfo {author} {\bibfnamefont {M.}~\bibnamefont {Gay~Ducati}},
  \ and\ \bibinfo {author} {\bibfnamefont {G.}~\bibnamefont {Soyez}},\ }\href
  {\doibase 10.1590/S0103-97332007000400032} {\bibfield  {journal} {\bibinfo
  {journal} {Braz. J. Phys.}\ }\textbf {\bibinfo {volume} {37}},\ \bibinfo
  {pages} {648} (\bibinfo {year} {2007}{\natexlab{a}})}\BibitemShut {NoStop}%
\bibitem [{\citenamefont {de~Santana~Amaral}\ \emph
  {et~al.}(2007{\natexlab{b}})\citenamefont {de~Santana~Amaral}, \citenamefont
  {Gay~Ducati}, \citenamefont {Betemps},\ and\ \citenamefont
  {Soyez}}]{deSantanaAmaral:2006fe}%
  \BibitemOpen
  \bibfield  {author} {\bibinfo {author} {\bibfnamefont {J.}~\bibnamefont
  {de~Santana~Amaral}}, \bibinfo {author} {\bibfnamefont {M.}~\bibnamefont
  {Gay~Ducati}}, \bibinfo {author} {\bibfnamefont {M.}~\bibnamefont {Betemps}},
  \ and\ \bibinfo {author} {\bibfnamefont {G.}~\bibnamefont {Soyez}},\ }\href
  {\doibase 10.1103/PhysRevD.76.094018} {\bibfield  {journal} {\bibinfo
  {journal} {Phys. Rev. D}\ }\textbf {\bibinfo {volume} {76}},\ \bibinfo
  {pages} {094018} (\bibinfo {year} {2007}{\natexlab{b}})},\ \Eprint
  {http://arxiv.org/abs/hep-ph/0612091} {arXiv:hep-ph/0612091} \BibitemShut
  {NoStop}%
\bibitem [{\citenamefont {Mueller}(1994)}]{Mueller:1993rr}%
  \BibitemOpen
  \bibfield  {author} {\bibinfo {author} {\bibfnamefont {A.~H.}\ \bibnamefont
  {Mueller}},\ }\href {\doibase 10.1016/0550-3213(94)90116-3} {\bibfield
  {journal} {\bibinfo  {journal} {Nucl. Phys. B}\ }\textbf {\bibinfo {volume}
  {415}},\ \bibinfo {pages} {373} (\bibinfo {year} {1994})}\BibitemShut
  {NoStop}%
\bibitem [{\citenamefont {Mueller}\ and\ \citenamefont
  {Patel}(1994)}]{Mueller:1994jq}%
  \BibitemOpen
  \bibfield  {author} {\bibinfo {author} {\bibfnamefont {A.~H.}\ \bibnamefont
  {Mueller}}\ and\ \bibinfo {author} {\bibfnamefont {B.}~\bibnamefont
  {Patel}},\ }\href {\doibase 10.1016/0550-3213(94)90284-4} {\bibfield
  {journal} {\bibinfo  {journal} {Nucl. Phys. B}\ }\textbf {\bibinfo {volume}
  {425}},\ \bibinfo {pages} {471} (\bibinfo {year} {1994})},\ \Eprint
  {http://arxiv.org/abs/hep-ph/9403256} {arXiv:hep-ph/9403256} \BibitemShut
  {NoStop}%
\bibitem [{\citenamefont {Mueller}(1995)}]{Mueller:1994gb}%
  \BibitemOpen
  \bibfield  {author} {\bibinfo {author} {\bibfnamefont {A.~H.}\ \bibnamefont
  {Mueller}},\ }\href {\doibase 10.1016/0550-3213(94)00480-3} {\bibfield
  {journal} {\bibinfo  {journal} {Nucl. Phys. B}\ }\textbf {\bibinfo {volume}
  {437}},\ \bibinfo {pages} {107} (\bibinfo {year} {1995})},\ \Eprint
  {http://arxiv.org/abs/hep-ph/9408245} {arXiv:hep-ph/9408245} \BibitemShut
  {NoStop}%
\bibitem [{\citenamefont {Yang}\ \emph {et~al.}(2020)\citenamefont {Yang},
  \citenamefont {Kou}, \citenamefont {Wang},\ and\ \citenamefont
  {Chen}}]{Yang:2020jmt}%
  \BibitemOpen
  \bibfield  {author} {\bibinfo {author} {\bibfnamefont {Y.}~\bibnamefont
  {Yang}}, \bibinfo {author} {\bibfnamefont {W.}~\bibnamefont {Kou}}, \bibinfo
  {author} {\bibfnamefont {X.}~\bibnamefont {Wang}}, \ and\ \bibinfo {author}
  {\bibfnamefont {X.}~\bibnamefont {Chen}},\ }\href@noop {} {\  (\bibinfo
  {year} {2020})},\ \Eprint {http://arxiv.org/abs/2009.11378} {arXiv:2009.11378
  [nlin.PS]} \BibitemShut {NoStop}%
\bibitem [{\citenamefont {Marquet}\ \emph {et~al.}(2005)\citenamefont
  {Marquet}, \citenamefont {Peschanski},\ and\ \citenamefont
  {Soyez}}]{Marquet:2005ic}%
  \BibitemOpen
  \bibfield  {author} {\bibinfo {author} {\bibfnamefont {C.}~\bibnamefont
  {Marquet}}, \bibinfo {author} {\bibfnamefont {R.~B.}\ \bibnamefont
  {Peschanski}}, \ and\ \bibinfo {author} {\bibfnamefont {G.}~\bibnamefont
  {Soyez}},\ }\href {\doibase 10.1016/j.physletb.2005.09.035} {\bibfield
  {journal} {\bibinfo  {journal} {Phys. Lett. B}\ }\textbf {\bibinfo {volume}
  {628}},\ \bibinfo {pages} {239} (\bibinfo {year} {2005})},\ \Eprint
  {http://arxiv.org/abs/hep-ph/0509074} {arXiv:hep-ph/0509074} \BibitemShut
  {NoStop}%
\bibitem [{\citenamefont {Golec-Biernat}\ and\ \citenamefont
  {Wusthoff}(1998)}]{GolecBiernat:1998js}%
  \BibitemOpen
  \bibfield  {author} {\bibinfo {author} {\bibfnamefont {K.~J.}\ \bibnamefont
  {Golec-Biernat}}\ and\ \bibinfo {author} {\bibfnamefont {M.}~\bibnamefont
  {Wusthoff}},\ }\href {\doibase 10.1103/PhysRevD.59.014017} {\bibfield
  {journal} {\bibinfo  {journal} {Phys. Rev.}\ }\textbf {\bibinfo {volume}
  {D59}},\ \bibinfo {pages} {014017} (\bibinfo {year} {1998})},\ \Eprint
  {http://arxiv.org/abs/hep-ph/9807513} {arXiv:hep-ph/9807513 [hep-ph]}
  \BibitemShut {NoStop}%
\bibitem [{\citenamefont {Bartels}\ \emph {et~al.}(2002)\citenamefont
  {Bartels}, \citenamefont {Golec-Biernat},\ and\ \citenamefont
  {Kowalski}}]{Bartels:2002cj}%
  \BibitemOpen
  \bibfield  {author} {\bibinfo {author} {\bibfnamefont {J.}~\bibnamefont
  {Bartels}}, \bibinfo {author} {\bibfnamefont {K.~J.}\ \bibnamefont
  {Golec-Biernat}}, \ and\ \bibinfo {author} {\bibfnamefont {H.}~\bibnamefont
  {Kowalski}},\ }\href {\doibase 10.1103/PhysRevD.66.014001} {\bibfield
  {journal} {\bibinfo  {journal} {Phys. Rev. D}\ }\textbf {\bibinfo {volume}
  {66}},\ \bibinfo {pages} {014001} (\bibinfo {year} {2002})},\ \Eprint
  {http://arxiv.org/abs/hep-ph/0203258} {arXiv:hep-ph/0203258} \BibitemShut
  {NoStop}%
\bibitem [{\citenamefont {Kowalski}\ and\ \citenamefont
  {Teaney}(2003)}]{Kowalski:2003hm}%
  \BibitemOpen
  \bibfield  {author} {\bibinfo {author} {\bibfnamefont {H.}~\bibnamefont
  {Kowalski}}\ and\ \bibinfo {author} {\bibfnamefont {D.}~\bibnamefont
  {Teaney}},\ }\href {\doibase 10.1103/PhysRevD.68.114005} {\bibfield
  {journal} {\bibinfo  {journal} {Phys. Rev. D}\ }\textbf {\bibinfo {volume}
  {68}},\ \bibinfo {pages} {114005} (\bibinfo {year} {2003})},\ \Eprint
  {http://arxiv.org/abs/hep-ph/0304189} {arXiv:hep-ph/0304189} \BibitemShut
  {NoStop}%
\bibitem [{\citenamefont {Breitweg}\ \emph {et~al.}(2000)\citenamefont
  {Breitweg} \emph {et~al.}}]{Breitweg:2000yn}%
  \BibitemOpen
  \bibfield  {author} {\bibinfo {author} {\bibfnamefont {J.}~\bibnamefont
  {Breitweg}} \emph {et~al.} (\bibinfo {collaboration} {ZEUS}),\ }\href
  {\doibase 10.1016/S0370-2693(00)00793-0} {\bibfield  {journal} {\bibinfo
  {journal} {Phys. Lett.}\ }\textbf {\bibinfo {volume} {B487}},\ \bibinfo
  {pages} {53} (\bibinfo {year} {2000})},\ \Eprint
  {http://arxiv.org/abs/hep-ex/0005018} {arXiv:hep-ex/0005018 [hep-ex]}
  \BibitemShut {NoStop}%
\bibitem [{\citenamefont {Chekanov}\ \emph {et~al.}(2001)\citenamefont
  {Chekanov} \emph {et~al.}}]{Chekanov:2001qu}%
  \BibitemOpen
  \bibfield  {author} {\bibinfo {author} {\bibfnamefont {S.}~\bibnamefont
  {Chekanov}} \emph {et~al.} (\bibinfo {collaboration} {ZEUS}),\ }\href
  {\doibase 10.1007/s100520100749} {\bibfield  {journal} {\bibinfo  {journal}
  {Eur. Phys. J.}\ }\textbf {\bibinfo {volume} {C21}},\ \bibinfo {pages} {443}
  (\bibinfo {year} {2001})},\ \Eprint {http://arxiv.org/abs/hep-ex/0105090}
  {arXiv:hep-ex/0105090 [hep-ex]} \BibitemShut {NoStop}%
\bibitem [{\citenamefont {Armesto}\ \emph {et~al.}(2005)\citenamefont
  {Armesto}, \citenamefont {Salgado},\ and\ \citenamefont
  {Wiedemann}}]{Armesto:2004ud}%
  \BibitemOpen
  \bibfield  {author} {\bibinfo {author} {\bibfnamefont {N.}~\bibnamefont
  {Armesto}}, \bibinfo {author} {\bibfnamefont {C.~A.}\ \bibnamefont
  {Salgado}}, \ and\ \bibinfo {author} {\bibfnamefont {U.~A.}\ \bibnamefont
  {Wiedemann}},\ }\href {\doibase 10.1103/PhysRevLett.94.022002} {\bibfield
  {journal} {\bibinfo  {journal} {Phys. Rev. Lett.}\ }\textbf {\bibinfo
  {volume} {94}},\ \bibinfo {pages} {022002} (\bibinfo {year} {2005})},\
  \Eprint {http://arxiv.org/abs/hep-ph/0407018} {arXiv:hep-ph/0407018 [hep-ph]}
  \BibitemShut {NoStop}%
\bibitem [{\citenamefont {Ben}\ \emph {et~al.}(2017)\citenamefont {Ben},
  \citenamefont {Machado},\ and\ \citenamefont {Sauter}}]{Ben:2017xny}%
  \BibitemOpen
  \bibfield  {author} {\bibinfo {author} {\bibfnamefont {F.~G.}\ \bibnamefont
  {Ben}}, \bibinfo {author} {\bibfnamefont {M.~V.~T.}\ \bibnamefont {Machado}},
  \ and\ \bibinfo {author} {\bibfnamefont {W.~K.}\ \bibnamefont {Sauter}},\
  }\href {\doibase 10.1103/PhysRevD.96.054015} {\bibfield  {journal} {\bibinfo
  {journal} {Phys. Rev. D}\ }\textbf {\bibinfo {volume} {96}},\ \bibinfo
  {pages} {054015} (\bibinfo {year} {2017})},\ \Eprint
  {http://arxiv.org/abs/1701.01141} {arXiv:1701.01141 [hep-ph]} \BibitemShut
  {NoStop}%
\bibitem [{\citenamefont {Golec-Biernat}\ and\ \citenamefont
  {Sapeta}(2018)}]{Golec-Biernat:2017lfv}%
  \BibitemOpen
  \bibfield  {author} {\bibinfo {author} {\bibfnamefont {K.}~\bibnamefont
  {Golec-Biernat}}\ and\ \bibinfo {author} {\bibfnamefont {S.}~\bibnamefont
  {Sapeta}},\ }\href {\doibase 10.1007/JHEP03(2018)102} {\bibfield  {journal}
  {\bibinfo  {journal} {JHEP}\ }\textbf {\bibinfo {volume} {03}},\ \bibinfo
  {pages} {102} (\bibinfo {year} {2018})},\ \Eprint
  {http://arxiv.org/abs/1711.11360} {arXiv:1711.11360 [hep-ph]} \BibitemShut
  {NoStop}%
\bibitem [{\citenamefont {Enberg}(2004)}]{Enberg:2004rz}%
  \BibitemOpen
  \bibfield  {author} {\bibinfo {author} {\bibfnamefont {R.}~\bibnamefont
  {Enberg}},\ }\href {\doibase 10.1142/S0217732304015956} {\bibfield  {journal}
  {\bibinfo  {journal} {Mod. Phys. Lett. A}\ }\textbf {\bibinfo {volume}
  {19}},\ \bibinfo {pages} {2655} (\bibinfo {year} {2004})},\ \Eprint
  {http://arxiv.org/abs/hep-ph/0410073} {arXiv:hep-ph/0410073} \BibitemShut
  {NoStop}%
\bibitem [{\citenamefont {Cai}\ \emph {et~al.}(2020)\citenamefont {Cai},
  \citenamefont {Xiang}, \citenamefont {Wang},\ and\ \citenamefont
  {Zhou}}]{Cai:2020exu}%
  \BibitemOpen
  \bibfield  {author} {\bibinfo {author} {\bibfnamefont {Y.}~\bibnamefont
  {Cai}}, \bibinfo {author} {\bibfnamefont {W.}~\bibnamefont {Xiang}}, \bibinfo
  {author} {\bibfnamefont {M.}~\bibnamefont {Wang}}, \ and\ \bibinfo {author}
  {\bibfnamefont {D.}~\bibnamefont {Zhou}},\ }\href {\doibase
  10.1088/1674-1137/44/7/074110} {\bibfield  {journal} {\bibinfo  {journal}
  {Chin. Phys. C}\ }\textbf {\bibinfo {volume} {44}},\ \bibinfo {pages}
  {074110} (\bibinfo {year} {2020})},\ \Eprint
  {http://arxiv.org/abs/2002.12610} {arXiv:2002.12610 [hep-ph]} \BibitemShut
  {NoStop}%
\bibitem [{\citenamefont {Accardi}\ \emph {et~al.}(2016)\citenamefont {Accardi}
  \emph {et~al.}}]{Accardi:2012qut}%
  \BibitemOpen
  \bibfield  {author} {\bibinfo {author} {\bibfnamefont {A.}~\bibnamefont
  {Accardi}} \emph {et~al.},\ }\href {\doibase 10.1140/epja/i2016-16268-9}
  {\bibfield  {journal} {\bibinfo  {journal} {Eur. Phys. J. A}\ }\textbf
  {\bibinfo {volume} {52}},\ \bibinfo {pages} {268} (\bibinfo {year} {2016})},\
  \Eprint {http://arxiv.org/abs/1212.1701} {arXiv:1212.1701 [nucl-ex]}
  \BibitemShut {NoStop}%
\bibitem [{\citenamefont {Chen}(2018)}]{Chen:2018wyz}%
  \BibitemOpen
  \bibfield  {author} {\bibinfo {author} {\bibfnamefont {X.}~\bibnamefont
  {Chen}},\ }\href {\doibase 10.22323/1.316.0170} {\bibfield  {journal}
  {\bibinfo  {journal} {PoS}\ }\textbf {\bibinfo {volume} {DIS2018}},\ \bibinfo
  {pages} {170} (\bibinfo {year} {2018})},\ \Eprint
  {http://arxiv.org/abs/1809.00448} {arXiv:1809.00448 [nucl-ex]} \BibitemShut
  {NoStop}%
\bibitem [{\citenamefont {Chen}\ \emph {et~al.}(2020)\citenamefont {Chen},
  \citenamefont {Guo}, \citenamefont {Roberts},\ and\ \citenamefont
  {Wang}}]{Chen:2020ijn}%
  \BibitemOpen
  \bibfield  {author} {\bibinfo {author} {\bibfnamefont {X.}~\bibnamefont
  {Chen}}, \bibinfo {author} {\bibfnamefont {F.-K.}\ \bibnamefont {Guo}},
  \bibinfo {author} {\bibfnamefont {C.~D.}\ \bibnamefont {Roberts}}, \ and\
  \bibinfo {author} {\bibfnamefont {R.}~\bibnamefont {Wang}},\ }\href {\doibase
  10.1007/s00601-020-01574-0} {\bibfield  {journal} {\bibinfo  {journal} {Few
  Body Syst.}\ }\textbf {\bibinfo {volume} {61}},\ \bibinfo {pages} {43}
  (\bibinfo {year} {2020})},\ \Eprint {http://arxiv.org/abs/2008.00102}
  {arXiv:2008.00102 [hep-ph]} \BibitemShut {NoStop}%
\end{thebibliography}%

\end{document}